\documentclass[pdflatex,sn-nature]{sn-jnl}

\usepackage{times}
\usepackage{graphicx}%
\usepackage{multirow}%
\usepackage{amsmath,amssymb,amsfonts,bm,braket}%
\usepackage{amsthm}%
\usepackage{mathrsfs}%
\usepackage[title]{appendix}%
\usepackage{xcolor}%
\usepackage{textcomp}%
\usepackage{manyfoot}%
\usepackage{booktabs}%
\usepackage{algorithm}%
\usepackage{algorithmicx}%
\usepackage{algpseudocode}%
\usepackage{listings}%
\usepackage{upgreek}

\author[1,2]{\fnm{Dasom} \sur{Kim}}
\author[3]{\fnm{Jin} \sur{Hou}}
\author[4,5]{\fnm{Geon} \sur{Lee}}
\author[6]{\fnm{Ayush} \sur{Agrawal}}
\author[7]{\fnm{Sunghwan} \sur{Kim}}
\author[1,6]{\fnm{Hao} \sur{Zhang}}
\author[8]{\fnm{Di} \sur{Bao}}
\author[2,9,10]{\fnm{Andrey} \sur{Baydin}}
\author[1,2]{\fnm{Wenjing} \sur{Wu}}
\author[1,2]{\fnm{Fuyang} \sur{Tay}}
\author[2,9]{\fnm{Shengxi} \sur{Huang}}
\author[8]{\fnm{Elbert} \sur{E.~M.~Chia}}
\author[5,7]{\fnm{Dai-Sik} \sur{Kim}}
\author[4,11]{\fnm{Minah} \sur{Seo}}
\author[3,6,9]{\fnm{Aditya D.} \sur{Mohite}}
\author[12]{\fnm{David} \sur{Hagenm\"uller}}
\author*[2,3,8,9,10,13]{\fnm{Junichiro} \sur{Kono}}\email{kono@rice.edu}

\affil[1]{\orgdiv{Applied Physics Graduate Program, Smalley--Curl Institute}, \orgname{Rice University}, \orgaddress{\city{Houston}, \state{Texas} \postcode{77005}, \country{USA}}}
\affil[2]{\orgdiv{Department of Electrical and Computer Engineering}, \orgname{Rice University}, \orgaddress{\city{Houston}, \state{Texas} \postcode{77005}, \country{USA}}}
\affil[3]{\orgdiv{Department of Materials Science and NanoEngineering}, \orgname{Rice University}, \orgaddress{\city{Houston}, \state{Texas} \postcode{77005}, \country{USA}}}
\affil[4]{\orgdiv{Sensor System Research Center}, \orgname{Korea Institute of Science and Technology}, \orgaddress{\city{Seoul}, \postcode{02792}, \country{Republic of Korea}}}
\affil[5]{\orgdiv{Department of Physics and Astronomy}, \orgname{Seoul National University}, \orgaddress{\city{Seoul}, \postcode{08826}, \country{Republic of Korea}}}
\affil[6]{\orgdiv{Department of Chemical and Biomolecular Engineering}, \orgname{Rice University}, \orgaddress{\city{Houston}, \postcode{77005}, \country{USA}}}
\affil[7]{\orgdiv{Department of Physics, Long-wavelength Nanotechnology Laboratory, and Quantum Photonics Institute}, \orgname{Ulsan National Institute of Science and Technology (UNIST)}, \orgaddress{\city{Ulsan}, \postcode{44919}, \country{Republic of Korea}}}
\affil[8]{\orgdiv{Division of Physics and Applied Physics, School of Physical and Mathematical Sciences}, \orgname{Nanyang Technological University}, \orgaddress{\city{Singapore}, \postcode{637371}, \country{Singapore}}}
\affil[9]{\orgdiv{Smalley--Curl Institute}, \orgname{Rice University}, \orgaddress{\city{Houston}, \state{Texas} \postcode{77005}, \country{USA}}}
\affil[10]{\orgdiv{Rice Advanced Materials Institute}, \orgname{Rice University}, \orgaddress{\city{Houston}, \state{Texas} \postcode{77005}, \country{USA}}}
\affil[11]{\orgdiv{KU–KIST Graduate School of Converging Science and Technology}, \orgname{Korea University}, \orgaddress{\city{Seoul}, \postcode{02841}, \country{Republic of Korea}}}
\affil[12]{\orgdiv{Institut de Physique et Chimie des Mat\'{e}riaux de Strasbourg (UMR 7504)}, \orgname{Universit\'e de Strasbourg and CNRS}, \orgaddress{\city{Strasbourg}, \postcode{67200}, \country{France}}}
\affil[13]{\orgdiv{Department of Physics and Astronomy}, \orgname{Rice University}, \orgaddress{\city{Houston}, \state{Texas} \postcode{77005}, \country{USA}}}

\begin{document}
\title{Multimode Phonon-Polaritons in the Ultrastrong Coupling Regime}

\maketitle


\section*{Abstract}

\textbf{Phonons play a central role in fundamental solid-state phenomena, including superconductivity, Raman scattering, and symmetry-breaking phases. Harnessing phonons to control these effects and enable quantum technologies is therefore of great interest. However, most existing phonon control strategies rely on external driving fields or anharmonic interactions, limiting their applicability. Here, we realize multimode ultrastrong light--matter coupling and theoretically show the modulation of phonon emission. This regime is realized by coupling two optical phonon modes in lead halide perovskites to a nanoslot array functioning as a single-mode cavity. The small mode volume of the nanoslots enables high coupling strengths in the phonon-polariton system. We show theoretically that the nanoslot resonator mediates an effective interaction between phonon modes, leading to superthermal phonon bunching in thermal equilibrium between distinct modes. Our findings are well described by a multimode Hopfield model. This work establishes a pathway for engineering phononic properties for light-harvesting and light-emitting technologies.}

\section*{Introduction}

Over the past few decades, metal halide perovskites have gained significant attention for potential use in solar cells~\cite{Kojima2009,Rong2018,Kim2020_2}. However, their carrier mobilities are generally lower than those of conventional inorganic semiconductors, largely due to strong electron--phonon interactions~\cite{Wright2016,Qu2023}. This has led to the growing interest in phonon engineering within perovskites, as it can profoundly influence carrier mobility, and, by extension, the energy conversion efficiency of devices.

The coherent manipulation of phonons using strong external laser fields has recently sparked considerable interest~\cite{Fausti2011,Mitrano2016,Kim2017,XLi2019,Zhou2020}. For instance, intense terahertz (THz) radiation can modify the band gap~\cite{Kim2017} and photoluminescence spectra~\cite{Sekiguchi2021} of perovskites. An alternative approach for controlling phonon properties involves cavity phonon-polaritons, which result from the coupling of phonons to the vacuum field of a cavity resonator~\cite{Kim2020,Yoo2021,Zhang2021,Baydin2023,Roh2023,Virgilio2023}. This method bypasses the need for external light sources and the challenges associated with phonon anharmonicities. By exploiting vacuum fluctuations, the phonon properties can be engineered by adjusting the resonator geometry~\cite{Fong2019}. Particularly, deep subwavelength cavities make it possible to observe phonon polaritons in nanoscale samples, with dimensions comparable to the carrier diffusion length, offering a promising strategy for mitigating carrier recombination in solar cells.

The coupling of phonons to THz subwavelength cavities also opens the door to the ultrastrong coupling (USC) regime of light--matter interaction~\cite{Forn-Díaz2019,Kockum2019}, where the coupling strength $g$ becomes comparable to the bare mode frequencies. In this regime, the counter-rotating terms in the Hamiltonian result in the ground state becoming a squeezed vacuum~\cite{Ciuti2005,Li2018}. Recent studies have shown that USC can give rise to remarkable phenomena, such as changes in electronic quantum transport~\cite{Appugliese2022}, tunable couplings between magnetic excitations~\cite{Makihara2021}, and magnonic superradiant phase transitions~\cite{Kim2024}. A particularly fascinating aspect of the USC regime is the appearance of anomalous correlations in the polaritonic ground state, which contains both photon and matter excitations~\cite{Ciuti2005}.

A recent study on the impact of phonon–photon coupling on electron--phonon interactions in perovskites, using ultrafast pump-probe spectroscopy, showed that the mobility of photoexcited carriers remained unaffected by light--matter coupling~\cite{Virgilio2023}. This study was conducted in the strong coupling regime, where the ground state is a standard vacuum. Investigating this behavior under USC conditions remains an intriguing prospect.

Recent works have reported the observation of a single phonon mode in lead halide perovskites coupled to a THz resonator~\cite{Kim2020,Zhang2021,Roh2023}. The possibility of coupling multiple phonon modes ultrastrongly to the resonator opens up new avenues for modifying electron--phonon interactions in the material. In recent years, multimode light--matter coupling has gained increasing attention across various platforms~\cite{Melnikau2019,Balasubrahmaniyam2021,He2022,Cortese2023,Tay2023,Mornhinweg2024}. Notably, multimode USC has been shown to induce ground-state correlations between different cavity modes~\cite{Tay2023}. However, the impact of multimode USC on intensity fluctuations in phonon-polariton systems has not yet been explored, despite the study of thermal photon statistics for a single two-level system in the USC regime~\cite{Ridolfo2013}.

In this work, we report the observation of multimode USC between two optical phonon modes, with frequencies $\omega_{1}$ and $\omega_{2}$, in 3D MAPbI$_3$ and 2D (BA)$_2$(MA)$_{1}$Pb$_2$I$_{7}$ lead halide perovskite crystals embedded in THz nanoslot cavities (Fig.~\ref{Fig1_main}a). By leveraging the extremely small mode volume of the nanoslots, which significantly enhances the coupling strengths $g_1$ and $g_2$ of each phonon mode, we achieved normalized coupling strengths at resonance of $g_{i}/\omega_{i} \sim 0.3$ ($i=1,2$). By tuning the nanoslot resonator frequency, we observed three distinct phonon-polaritons, which exhibited two Rabi splittings in THz time-domain spectroscopy (THz-TDS) measurements. These experimental results were consistent with numerical simulations and calculations based on a Hopfield quantum model. In a polaritonic thermal state at room temperature, the model predicts the presence of ``superthermal'' phonon bunching in the off-resonance regime, where the nanoslot resonator frequency is much smaller than the bare phonon frequencies. For a resonator frequency $\omega_\mathrm{c}/(2\pi)=0.1$~THz, the intramode equal-time second-order correlation function $g^{(2)}_{ii}(\tau=0)$, which quantifies the probability of simultaneous phonon emission in mode $i=1,2$, exceeds the value $g^{(2)}_{ii}(0) = 2$ for bare phonons at thermal equilibrium and is found to be governed by the USC figure of merit $g_{i}/\omega_{i}$. Moreover, while phonon emission in two distinct modes is uncorrelated without light--matter coupling (i.e., $g^{(2)}_{12}(0) = 1$), we show that multimode USC results in pronounced intermode bunching ($g^{(2)}_{12}(0) \approx 3$), governed by the figure of merit $g_{1}g_{2}/\omega_{1}\omega_{2}$.

\section*{Experimental results}

We fabricated an array of nanoslots ($w=950$~nm) on quartz substrates with seven different lengths ($l =$ 30, 40, 50, 60, 80, 120, and 160~$\upmu$m) to tune the cavity mode frequency, given by $\omega_\mathrm{c}/(2\pi)=c_{0}/(2l\sqrt{\epsilon_{\textrm{avg}}})$, where $c_0$ is the speed of light in vacuum and $\epsilon_{\textrm{avg}}=(\epsilon_{\textrm{air}}+\epsilon_{\textrm{sub}})/2$ represents the average dielectric constant of air and the quartz substrate ($\epsilon_{\textrm{sub}}=2.1^2$)~\cite{Kang2009}; see Methods for sample preparation details. The resonance frequency is predominantly governed by the geometry of a single nanoslot rather than that of the periodic array~\cite{Garcia-Vidal2005,Alù2008}. Figure~\ref{Fig1_main}b illustrates the structure of our samples, where perovskite films (purple) are coated both on top of and within the slots.

These films exhibit two distinct optical phonon modes in free space, labeled as $\lambda=1$ and $\lambda=2$, corresponding to the rocking and stretching of Pb--I bonds, respectively. Due to the orientational disorder of methylammonium molecules, which breaks the lattice space-group symmetry, these phonons acquire a mixed transverse-optical (TO) and longitudinal-optical (LO) character~\cite{La-o-vorakiat2016}. As a result, they not only exhibit strong infrared absorption~\cite{La-o-vorakiat2016,Sendner2016} but also interact with lattice electrons, as recently observed~\cite{Qu2023}. Moreover, low-frequency phonons are particularly beneficial for achieving ultra-strong coupling, as the normalized coupling strength $g/\omega$ increases with decreasing phonon frequency. This study, therefore, focuses on the phonons that are most relevant to strong interactions with both photonic and electronic degrees of freedom. Since electron-phonon interactions dictate charge mobility and recombination through long-range Coulomb forces, phonon-polariton formation involving low-frequency hybrid TO/LO phonons could offer novel pathways to engineer charge transport in lead-halide perovskites.

Nanoslot resonators provide significant electric field enhancement due to strong optical confinement within and around the slots~\cite{Seo2009,Kim2018}. Since the phonon–photon coupling strength $g \propto \sqrt{N/V}$, where $N$ is the number of unit cells in the crystal and $V$ is the resonator mode volume, the small mode volume of nanoslot resonators enables extreme light--matter interaction regimes even with small perovskite crystals.

The in-plane spatial distribution of the cavity mode, computed using COMSOL for a perovskite-filled nanoslot, is shown in Fig.~\ref{Fig1_main}c (left panel). The field profile follows a sinusoidal pattern~\cite{Roh2023} along the $y$-axis, with an electric field enhancement factor of 20 relative to transmission through a bare quartz substrate. The strong confinement of the $x$-component of the electric field ($E_x$) along the $x$- and $z$-axes results in a nearly uniform electric field within the perovskite region (Fig.~\ref{Fig1_main}c, right panel). Although the nanoslot thickness is 130\,nm, the cavity mode extends beyond the nanoslot into the surrounding MAPbI$_3$ layer, as depicted in Fig.~\ref{Fig1_main}c. The electric field strength above the nanoslot remains comparable to that inside, indicating that the perovskite film covering the slot also contributes to light--matter coupling. Notably, when $t$ becomes comparable to the mode's spatial extent along $z$, where $t$ is the perovskite film thickness, $g$ saturates at its maximum value (see Supplementary Information).

We characterized the perovskite–nanoslot hybrid system using THz-TDS at room temperature. A normal-incident THz beam was linearly polarized along the $x$-axis. In free space, a 200-nm-thick MAPbI$_3$ film exhibits transmittance dips at $\omega_{1}/(2\pi)=0.96$~THz and $\omega_{2}/(2\pi)=1.9$~THz, corresponding to the two phonon modes $\lambda=1$ and $\lambda=2$, respectively (Fig.~\ref{Fig2_main}a). The bare cavity resonance appears as a single peak in the transmission spectrum. Figure \ref{Fig2_main}b displays the cavity resonance frequency as a function of cavity length $l$. By adjusting $l$, the cavity mode can be brought into resonance with either the $\lambda=1$ mode ($\omega_\mathrm{c}=\omega_{1}$) or the $\lambda=2$ mode ($\omega_\mathrm{c} = \omega_{2}$).

Figure~\ref{Fig2_main}c presents the transmission spectra of MAPbI$_3$-–nanoslot structures for different $l$ values. As the nanoslots predominantly reflect incoming radiation, the observed polariton modes appear as transmission peaks. The spectra exhibit three distinct polariton branches: lower (LP), middle (MP), and upper (UP). These branches are separated by the uncoupled phonon modes $\lambda=1$ and $\lambda=2$ (dashed lines). As $l$ decreases, the LP branch shifts toward $\lambda=1$, the MP branch moves away from $\lambda=1$ and approaches $\lambda=2$, while the UP branch shifts away from $\lambda=2$. Two anticrossings are observed at $l=80\,\upmu$m and $l=50\,\upmu$m, corresponding to $\omega_\mathrm{c} \approx \omega_{1}$ and $\omega_\mathrm{c} \approx \omega_{2}$, respectively (Fig.~\ref{Fig2_main}b). Due to the larger oscillator strength of the $\lambda=2$ mode (Fig.~\ref{Fig2_main}a), the second Rabi splitting at $l=50\,\upmu$m exceeds the first at $l=80\,\upmu$m.

We carried out finite element simulations (COMSOL) to validate our experimental results, using conductivity values extracted from THz-TDS measurements (Fig.~S1) as input parameters. The simulated transmission spectra (Fig.~\ref{Fig2_main}d, colormap) closely match the experimental data, with black solid circles marking the resonance frequencies obtained from Fig.~\ref{Fig2_main}c via Lorentzian fitting. Minor discrepancies in the UP frequencies are attributed to slight shifts in the bare cavity mode (dashed green line, Fig.~\ref{Fig2_main}a) and additional coupling with a 3.8 THz phonon mode in the $z$-cut quartz substrate.

We also investigated a 2D perovskite material composed of metal-halide layers separated by organic molecules, which enhances stability compared to 3D perovskites~\cite{Smith2014,Tsai2016,Sidhik2024} and holds promise for solar cell applications. Unlike 3D MAPbI$_3$ (Fig.~\ref{Fig3_main}a), the presence of BA cations (CH$_3$(CH$_2$)$_3$NH$_3$) reduces the number of Pb–I bonds per unit volume, weakening the phonon mode oscillator strength. The layered structure, (BA)$_2$(MA)$_{n-1}$Pb$_n$I$_{3n+1}$ (with n=2)~\cite{Tsai2016}, is shown in Fig.~\ref{Fig3_main}b. Here, $n$ denotes the number of PbI$_6$ octahedral layers between the BA spacer layers. The phonon modes $\lambda=1$ and $\lambda=2$ are slightly blueshifted compared to MAPbI$_3$, with dips in the transmittance of a bare (BA)$_2$(MA)$_{1}$Pb$_2$I$_{7}$ 200-nm-thick film at $\omega_{1}/(2\pi)=1.09$~THz and $\omega_{2}/(2\pi)=2$~THz, respectively (Fig.~S2). The transmission spectra of 2D perovskites embedded in nanoslot resonators resemble those of their 3D counterparts, with a larger Rabi splitting for $\lambda=2$ due to its higher oscillator strength.

\section*{Theoretical analysis: Quantum model}

While classical electrodynamics simulations accurately reproduce the transmission spectra, we now adopt a complementary approach by utilizing a multimode Hopfield model~\cite{Hopfield1958} to gain a deeper understanding of the ultrastrong light–matter coupling in our system and investigate its potential implications. The microscopic Hamiltonian, derived in Section 2 of the Supplementary Information, is given by
\begin{align}
\hat{H}=\hbar \omega_\mathrm{c} \hat{a}^{\dagger} \hat{a} + \sum_{\lambda} \hbar \omega_{\lambda} \hat{b}^{\dagger}_{\lambda} \hat{b}_{\lambda} -i \sum_{\lambda} \hbar g_{\lambda} \left(\hat{b}^{\dagger}_{\lambda}-\hat{b}_{\lambda}\right) \left(\hat{a}+\hat{a}^{\dagger}\right) + \sum_{\lambda} \frac{\hbar g^{2}_{\lambda}}{\omega_{\lambda}} \left(\hat{a}+\hat{a}^{\dagger}\right)^{2},
\label{total_H}
\end{align}
where $\hat{a}^\dagger$ ($\hat{a}$) represents the creation (annihilation) operator of a cavity photon, while $\hat{b}^\dagger_{\lambda}$ ($\hat{b}_{\lambda}$) denotes the creation (annihilation) operator of a phonon in the mode $\lambda$. The first two terms correspond to the bare photon and phonon Hamiltonians, respectively. The third term describes the light--matter interaction, with a coupling strength given by $g_{\lambda}=\frac{\nu_{\lambda}}{2} \sqrt{\frac{\omega_{\lambda}}{\omega_\mathrm{c}}}$, which is proportional to the effective ion plasma frequency $\nu_{\lambda}$. The fourth term, known as the $A^2$-term, induces a blueshift in the cavity mode frequency. The effective ion plasma frequency is determined by the effective charges associated with Pb$^{2+}$ and I$^-$ ions, as detailed in Sec.~2 of the Supplementary Information.

The eigenfrequencies and eigenvectors of the Hamiltonian Eq.~\eqref{total_H} are obtained via the Hopfield transformation: $\hat{p}_{\alpha}=\sum_{\lambda} X_{\lambda,\alpha} \hat{b}_{\lambda}+\sum_{\lambda} \widetilde{X}_{\lambda,\alpha} \hat{b}^{\dagger}_{\lambda}+ Y_{\alpha} \hat{a} + \widetilde{Y}_{\alpha} \hat{a}^{\dagger}$, where $\hat{p}_{\alpha}$ is the annihilation operator of a polariton in the mode $\alpha=$ LP, MP, UP, with frequency $\omega_{\alpha}$. Up to a constant term, Eq.~\eqref{total_H} can then be expressed in its diagonal form as $\hat{H}=\sum_{\alpha}\hbar \omega_{\alpha} \hat{p}^{\dagger}_{\alpha}\hat{p}_{\alpha}$. The system enters the USC regime when the normalized coupling strength at resonance satisfies $g_{\lambda}/\omega_{\lambda}=\nu_{\lambda}/2\omega_{\lambda} \gtrsim 0.1$. In this regime, the counter-rotating terms $\propto \hat{b}_{\lambda}\hat{a}, \hat{b}^{\dagger}_{\lambda} \hat{a}^{\dagger}$ in the Hamiltonian Eq.~\eqref{total_H}, along with the anomalous Hopfield coefficients $\widetilde{Y}_{\alpha}$ and $\widetilde{X}_{\lambda,\alpha}$, play a significant role. 

Figure~\ref{Fig3_main}c presents the polariton dispersion for the MAPbI$_3$--nanoslots system. The coupling strengths $g_{\lambda}$ are extracted by fitting the peak frequencies (solid circles) of the transmission spectra to the calculated eigenfrequencies $\omega_{\alpha}$ (solid lines). When the nanoslot resonator is resonant with the phonon modes $\lambda=1$ and $\lambda=2$, we obtain normalized coupling strengths of $g_1/\omega_1=0.28$ ($\omega_\mathrm{c}=\omega_{1}$) and $g_2/\omega_2=0.3$ ($\omega_\mathrm{c}=\omega_{2}$), respectively. These values confirm that both phonon modes are in the USC regime with the nanoslot resonator. The corresponding Rabi splittings at the two resonances are $0.45$ THz and $1.13$ THz. Notably, while the Rabi splitting equals exactly $2g$ for a single matter and cavity mode in the strong coupling regime, this relation breaks down in the USC regime due to counter-rotating terms. In our case, the inclusion of two phonon modes leads to further deviations from the $2g$ value. Importantly, the polariton dispersion should be understood as the result of the simultaneous coupling of both phonon modes to the cavity mode, with all three degrees of freedom treated on equal footing. This becomes evident when examining the contribution of the two phonon modes to the MP mode at around the resonance between the $\lambda = 1$ phonon and the cavity mode. As shown in Fig.~S4b of the supplementary information, the contribution $W^{\textrm{MP}}_{\lambda} = \vert X_{\lambda,\textrm{MP}} \vert^{2} - \vert \widetilde{X}_{\lambda,\textrm{MP}} \vert^{2}$ of the two phonon modes ($\lambda = 1,2$) to the MP mode is indeed of comparable magnitude, indicating that the MP branch involves significant hybridization with both phonon modes.

In the USC regime, distinctive features appear not only near resonance, but also when the resonator frequency is much lower than the phonon frequencies, $\omega_\mathrm{c} \ll \omega_{\lambda}$. Unlike in the strong coupling regime, the polariton modes do not converge to the uncoupled mode frequencies when $\omega_\mathrm{c} \ll \omega_{\lambda}$. This behavior defines the so-called polariton gaps~\cite{Maissen2014UltrastrongResonators, Forn-Díaz2019}, given by $\Delta_{1} = \lim_{\omega_\mathrm{c} \to 0} \omega_{\textrm{MP}} - \omega_{1}$ and $\Delta_{2} = \lim_{\omega_\mathrm{c} \to 0} \omega_{\textrm{UP}} - \omega_{2}$, as shown in Fig.\ref{Fig3_main}c. As detailed in Sec.~3 of the Supplementary Information, the frequencies of the MP and UP modes, $\omega_{\textrm{MP}}$ and $\omega_{\textrm{UP}}$, asymptotically approach $\widetilde{\omega}_{1}=\sqrt{\omega_{1}^{2} + \nu_{1}^{2}}$ and $\widetilde{\omega}_{2}=\sqrt{\omega_{2}^{2} + \nu_{2}^{2}}$, respectively, in the low-resonator frequency limit $\omega_\mathrm{c} \to 0$. 

This unconventional behavior is linked to strong light--matter hybridization, which persists even in this far-detuned, low-resonator-frequency regime. This is reflected in the divergence of the light--matter coupling strength, $g_{\lambda} \propto \sqrt{1/\omega_\mathrm{c}}$, and the $A^2$ term, which scales as $g_{\lambda}^2$, as $\omega_\mathrm{c} \to 0$. In this regime, the MP (UP) mode is mainly a hybrid of the phonon mode $\lambda = 1$ ($\lambda = 2$) and cavity photons. The corresponding normal ($Y_{\alpha}$) and anomalous ($\widetilde{Y}_{\alpha}$) Hopfield coefficients become large and comparable, scaling as $Y_{\textrm{MP}} \sim \widetilde{Y}_{\textrm{MP}} \sim \nu_{1}/\sqrt{\omega_\mathrm{c}\widetilde{\omega}_{1}}$ and $Y_{\textrm{UP}} \sim \widetilde{Y}_{\textrm{UP}} \sim \nu_{2}/\sqrt{\omega_\mathrm{c}\widetilde{\omega}_{2}}$. In contrast, the LP mode mixes the cavity field with both phonon modes. The phonon contributions to this polariton, quantified by the coefficients $X_{\lambda,\textrm{LP}}$ and $\widetilde{X}_{\lambda,\textrm{LP}}$, also grow large and comparable, with $X_{\lambda,\textrm{LP}} \sim \widetilde{X}_{\lambda,\textrm{LP}} \sim \nu_{\lambda}/\sqrt{\omega_\mathrm{c}\omega_{\lambda}}$. These coefficients are shown in Fig.~\ref{Fig3_main}d, while the other Hopfield coefficients are provided in Figs.~S3 and S5.


Due to the large anomalous Hopfield coefficients, the polaritonic ground state $\ket{G}$, defined by $\prod_{\alpha}\hat{p}_{\alpha} \ket{G}=0$, takes the form of a multimode squeezed vacuum in the low resonator frequency regime. This state contains correlated photon pairs, contributed by the MP and UP, as well as intermode and intramode phonon pairs originating from the LP. This multimode squeezed vacuum exhibits strong entanglement between the two phonon modes, as discussed below.

It is important to highlight that while both the normal and anomalous Hopfield coefficients diverge in the limit $\omega_\mathrm{c} \to 0$, the total phonon and photon weights remain finite due to the normalization of the Hopfield coefficients (see Figs.~S4 and S6). Moreover, we stress that the low-resonator-frequency regime (long cavity) does not imply the absence of a cavity, as its transverse confinement remains deeply subwavelength.
  
A distinctive feature of the multimode USC regime is the presence of anomalous correlations between the phonon modes. By inverting the Hopfield transformation, one obtains the correlation functions: 
\begin{align}
\langle \hat{b}^{\dagger}_{\lambda} \hat{b}_{\lambda'} \rangle &= \sum_{\alpha} \left(\widetilde{X}^{\alpha}_{\lambda} \right)^{*} \widetilde{X}^{\alpha}_{\lambda'} (1+n_{\alpha}) + \sum_{\alpha} X^{\alpha}_{\lambda} \left(X^{\alpha}_{\lambda'} \right)^{*} n_{\alpha}, \nonumber \\
\langle \hat{b}_{\lambda} \hat{b}_{\lambda'} \rangle &= -\sum_{\alpha} \left( X^{\alpha}_{\lambda} \right)^{*} \widetilde{X}^{\alpha}_{\lambda'} (1+n_{\alpha}) - \sum_{\alpha} \left( X^{\alpha}_{\lambda'} \right)^{*} \widetilde{X}^{\alpha}_{\lambda} n_{\alpha}.
\label{correlations}
\end{align}
Here, $n_{\alpha}=\langle \hat{p}^{\dagger}_{\alpha}\hat{p}_{\alpha} \rangle$ represents the population in the polariton mode $\alpha$. In the polaritonic ground state ($n_{\alpha}=0$), Eq.~\eqref{correlations} shows that such correlations arise only when the anomalous Hopfield coefficients $\widetilde{X}^{\alpha}_{\lambda}$ are nonzero. Moreover, these correlations are further enhanced in excited polariton states where $n_{\alpha} \neq 0$. To explore the impact of multimode USC on correlated phonon emission, we consider the second-order correlation function~\cite{Humphries2023}, which quantifies the joint probability of a phonon being emitted in the mode $\lambda'$ at time $t+\tau$ given that a phonon was emitted in the mode $\lambda$ at time $t$:
\begin{align*}
g^{(2)}_{\lambda,\lambda'} (\tau)= \frac{\langle \hat{b}_{\lambda'} (t+\tau) \hat{b}_{\lambda} (t) \hat{b}^{\dagger}_{\lambda} (t)\hat{b}^{\dagger}_{\lambda'} (t+\tau)\rangle}{\langle \hat{b}_{\lambda} (t) \hat{b}^{\dagger}_{\lambda} (t) \rangle \langle \hat{b}_{\lambda'} (t+\tau) \hat{b}^{\dagger}_{\lambda'} (t+\tau) \rangle}.
\end{align*}
Although classical wave-based methods could, in principle, be used to study intensity correlations, such calculations are technically challenging and non-trivial to implement. For this reason, we focus in this work on quantum predictions for the second-order correlation functions. Assuming a thermal polariton state at temperature $T$, with $n_{\alpha}=\left(e^{\hbar \omega_{\alpha}/k_{\textrm{B}} T} -1 \right)^{-1}$ and $k_{\textrm{B}}$ the Boltzmann constant, the equal-time intramode ($\lambda=\lambda'$) and intermode ($\lambda\neq \lambda'$) correlation functions are given by
\begin{align}
g^{(2)}_{\lambda,\lambda} (0)&= 2 + \frac{\langle \hat{b}_{\lambda} \hat{b}_{\lambda} \rangle \langle \hat{b}^{\dagger}_{\lambda}\hat{b}^{\dagger}_{\lambda}\rangle}{\langle \hat{b}_{\lambda} \hat{b}^{\dagger}_{\lambda} \rangle^{2}}, \nonumber \\
g^{(2)}_{\lambda,\lambda'} (0) &= 1+\frac{\langle \hat{b}_{\lambda} \hat{b}^{\dagger}_{\lambda'} \rangle \langle \hat{b}_{\lambda'} \hat{b}^{\dagger}_{\lambda} \rangle}{\langle \hat{b}_{\lambda} \hat{b}^{\dagger}_{\lambda} \rangle \langle \hat{b}_{\lambda'} \hat{b}^{\dagger}_{\lambda'} \rangle} + \frac{\langle \hat{b}_{\lambda} \hat{b}_{\lambda'} \rangle \langle \hat{b}^{\dagger}_{\lambda'} \hat{b}^{\dagger}_{\lambda} \rangle}{\langle \hat{b}_{\lambda} \hat{b}^{\dagger}_{\lambda} \rangle \langle \hat{b}_{\lambda'} \hat{b}^{\dagger}_{\lambda'} \rangle}, 
\label{results_g2}
\end{align}
respectively. For vanishing anomalous Hopfield coefficients ($\widetilde{X}^{\alpha}_{\lambda}=0$ $\forall \lambda$) or in the absence of phonon--photon coupling ($\nu_{\lambda}=0$), Eq.~\eqref{results_g2} simplifies to $g^{(2)}_{\lambda,\lambda} (0)=2$, which corresponds to intramode phonon bunching—a hallmark of thermal states. In contrast, the intermode correlation function satisfies $g^{(2)}_{\lambda,\lambda'} (0)=1$ for $\lambda\neq \lambda'$, indicating that intermode phonon emission remains uncorrelated and follows Poissonian statistics. 

In the multimode USC regime, we predict a significant modification of equal-time phonon–phonon second-order correlations. As shown in Fig.~\ref{Fig3_main}e, the various contributions $g^{(2)}_{\lambda,\lambda'} (0)$ to a value of $3$ in the limit of a vanishing resonator frequency ($\omega_\mathrm{c} \to 0$). As $\omega_\mathrm{c}$ increases, $g^{(2)}_{\lambda,\lambda'} (0)$ decreases monotonically, approaching $2$ for intramode correlations ($\lambda=\lambda'$) and $1$ for intermode correlations ($\lambda\neq \lambda'$). These limiting values correspond to the correlations expected for bare phonons, which are recovered in the high resonator frequency regime ($\omega_\mathrm{c} \gg \omega_{\lambda}$), where the LP and MP asymptotically approach the uncoupled phonon frequencies $\omega_1$ and $\omega_2$, respectively.

For a detuned cavity with $\omega_\mathrm{c}/(2\pi)=0.1$~THz at room temperature ($T=300$~K), our theoretical model predicts $g^{(2)}_{1,1} (0)\approx 2.86$, $g^{(2)}_{2,2} (0)\approx 2.96$, and $g^{(2)}_{1,2} (0)\approx 2.82$. These results indicate that multimode USC should lead to strong phonon bunching in both intramode and intermode correlations. This effect primarily arises from the LP, which exhibits large normal and anomalous phonon Hopfield coefficients ($X^{\alpha}_{\lambda},\widetilde{X}^{\alpha}_{\lambda}$), as discussed earlier, along with a significant population in the low resonator frequency regime ($n_{\textrm{LP}}\approx 80$ for $\omega_\mathrm{c}/(2\pi)=0.1$~THz and $T=300$~K). The inset of Fig.~\ref{Fig3_main}e illustrates the temperature dependence of the calculated second-order phonon correlations, showing that at $T=0$~K, intramode and intermode correlations are enhanced by approximately 10\% and 40\%, respectively, compared to the bare phonon case. Notably, at room temperature, $g^{(2)}_{\lambda,\lambda'} (0)$ remains in the saturation regime. Note that the system also exhibits strong photon bunching (not shown), driven by correlated photon pairs primarily originating from the MP and UP contributions to the polaritonic ground state.

Figure~\ref{Fig3_main}f presents the extracted peak frequencies for the 2D perovskite (BA)$_2$(MA)$_{1}$Pb$_2$I$_{7}$--nanoslots system, alongside theoretical predictions (solid lines). In this case, the $\lambda=1$ mode exhibits a very small polaritonic gap, which is consistent with the normalized coupling strength $g'_{1}/\omega_{1} = 0.13$ extracted from the fit. This value suggests that $\lambda=1$ is on the verge of the USC regime, leading to reduced Hopfield coefficients ${X}_{\textrm{1,LP}}$ and $\widetilde{X}_{\textrm{1,LP}}$ compared to the MAPbI$_3$--nanoslots system (see Fig.~\ref{Fig3_main}g). Conversely, the polaritonic gap of the $\lambda=2$ mode remains similar to that observed in the MAPbI$_3$--nanoslots system, consistent with the large coupling ratio $g'_{2}/\omega_{2} = 0.23$ extracted from the fit.

At $\omega_\mathrm{c}/(2\pi) = 0.1$~THz and room temperature, the calculated phonon–phonon correlations for $\lambda=\lambda'=1$ and $\lambda=1,\lambda'=2$ are significantly lower than in the MAPbI$_3$–nanoslots system ($g^{(2)}_{1,1} (0)\approx 2.61$, $g^{(2)}_{1,2} (0)\approx 2.52$), in line with the weaker coupling strength of $\lambda=1$. However, $g^{(2)}_{2,2} (0)\approx 2.94$ remains nearly unchanged compared to the 3D perovskite system, as shown in Fig.\ref{Fig3_main}h. 

Using a perturbative expansion valid for $\omega_\mathrm{c}/\omega_{\lambda} \ll 1$ and $\nu_{\lambda}/\omega_{\lambda} \ll 1$, we show in Section 3 of Supplementary Information that the second-order correlation functions can be approximated as
\begin{align*}
g^{(2)}_{1,1} (0) & \approx 2 + \left(\frac{g_{1}}{\omega_{1}} \right)^{4} \left(\frac{1+2 n_{\textrm{LP}}}{1+n_{\textrm{MP}}} \right)^{2} \nonumber \\
g^{(2)}_{2,2} (0) & \approx 2 + \left(\frac{g_{2}}{\omega_{2}} \right)^{4} \left(\frac{1+2 n_{\textrm{LP}}}{1+n_{\textrm{UP}}} \right)^{2} \\
g^{(2)}_{1,2} (0) & \approx 1 + 2\left(\frac{g_{1}}{\omega_{1}} \right)^{2}\left(\frac{g_{2}}{\omega_{2}} \right)^{2}\frac{(1+2 n_{\textrm{LP}})^{2}}{(1+n_{\textrm{MP}})(1+n_{\textrm{UP}})}.
\end{align*}
These results show that the intramode correlation functions are primarily controlled by the standard USC figure of merit, $g_{\lambda}/\omega_{\lambda}$. In contrast, intermode correlations are governed by the product  $g_{1}g_{2}/\omega_{1}\omega_{2}$, which becomes an important figure of merit for multimode USC. For instance, we find $g_{1}g_{2}/\omega_{1}\omega_{2}=0.084$ in the 3D perovskite–nanoslot system and $g_{1}g_{2}/\omega_{1}\omega_{2} =0.03$ in the 2D system. The specific form $g_{1}g_{2}/\omega_{1}\omega_{2}$ suggests that intermode correlations arise from the effective coupling between phonons mediated by the far-detuned cavity, where $\omega_\mathrm{c}\ll \omega_{\lambda}$.  

\section*{Discussion}

We report the first observation of cavity phonon-polaritons in the multimode USC regime. The light--matter coupling strength was controlled by tuning the number of PbI$_6$ octahedral layers between the BA spacer layers of the perovskite, directly influencing the phonon oscillator strengths. Unlike recent studies on multimode USC, which have primarily focused on engineering photonic properties in THz cavities through coupling with inorganic semiconductor (GaAs) quantum wells~\cite{Tay2023,Cortese2023,Mornhinweg2024}, our complementary approach leverages a deep-subwavelength cavity resonator to mediate effective coupling between matter excitations. This approach has the potential to modify fundamental material properties, such as charge carrier mobilities. Given the relevance to solar cell applications, we focus on multimode USC of phonons in lead halide perovskite thin films, which are known to exhibit strong electron--phonon interactions~\cite{Gong2018,Yamada2022,Qu2023}. Our cavity-mediated phonon--phonon coupling mechanism provides a novel route for controlling phonon--phonon correlations at thermal equilibrium, without requiring external driving fields or phonon anharmonicities.

The exceptionally small mode volume of the nanoslots enabled USC with the highest resonant coupling strengths reported in cavity phonon-polariton systems. The use of deep-subwavelength resonators filled with lead halide perovskite films of a few hundred nanometers in thickness---comparable to the carrier diffusion length---is fully compatible with solar cell applications~\cite{Momblona2014}. In the off-resonance regime, where the cavity frequency is much lower than the phonon frequencies, the coupling strength scales as $g_{\lambda} \propto 1/\sqrt{\omega_\mathrm{c}}$, allowing access to a unique regime where counter-rotating terms in the Hamiltonian become as significant as the rotating-wave terms. This leads to anomalous correlations governed by the USC figure of merit $g_{\lambda}/\omega_{\lambda}$ at thermal equilibrium, even in the absence of nonlinear interactions. We demonstrate theoretically that in this regime, the cavity mode mediates an effective interaction between the two phonon modes $\lambda$ and $\lambda'$, resulting in superthermal intermode phonon bunching $\propto g_{\lambda}g_{\lambda'}/\sqrt{\omega_{\lambda}\omega_{\lambda'}}$. This corresponds to the correlated emission of phonons, characterized by an equal-time second-order phonon--phonon correlation function $g^{(2)}_{\lambda,\lambda'} (0) > 2$. In contrast, for bare phonons in thin films without a cavity, phonon emission in different modes remains uncorrelated, i.e., $g^{(2)}_{\lambda,\lambda'} (0) = 1$. 

Although directly measuring phonon--phonon correlations remains challenging, indirect evidence can be obtained through quantum optics techniques that measure the second-order photon--photon correlation function. As mentioned earlier, this function also exhibits superthermal bunching and can be directly measured using femtosecond noise correlation spectroscopy, a method previously used to study mode fluctuations in quantum fields~\cite{Benea-Chelmus2019ElectricState}. Additionally, a recent study on magnons demonstrated that statistical correlations between two probe pulses can be used to extract fluctuations of collective excitations~\cite{Weiss2023}, a technique that can be adapted to phonons. Finally, nonlinear spectroscopic techniques, such as two-dimensional THz spectroscopy, can provide further indirect insights into phonon--phonon correlations. 

Compared to recent studies on single-mode phonon-polaritons in similar systems~\cite{Kim2020,Zhang2021,Roh2023}, our multimode scenario presents new opportunities for controlling electron--phonon interactions in lead halide perovskites, with implications for light-harvesting and light-emitting devices. This advantage stems from the unique nature of the phonons studied here: they are strongly coupled both to the cavity and to charge carriers due to their mixed TO/LO character. Furthermore, we show that by employing sufficiently long resonators, it is possible to achieve a high phonon-to-cavity frequency ratio, $\omega/\omega_\mathrm{c}$, which effectively compensates for the small coupling ratio, $\nu/\omega$, of higher-energy phonons with a predominantly LO character and consequently low oscillator strength. Previous studies have demonstrated that charge carrier scattering, mediated by the Fr\"{o}hlich interaction with LO phonons near 3 THz, dominates electron--phonon coupling in these materials at room temperature~\cite{Wright2016}. In long nanoslots, this compensation effect would enhance the coupling strength of these LO phonons, $g/\omega=(\nu/\omega) \sqrt{\omega/\omega_\mathrm{c}}$, potentially yielding a large multimode USC figure of merit and, therefore, strong intermode phonon bunching with the low-frequency phonons $\lambda=1,2$.

Our approach thus enables control over high-frequency LO phonons via USC coupling to low-frequency IR-active phonons in long cavities, potentially leading to substantial modifications in electron--phonon scattering. This motivates further pump-probe photoconductivity experiments under multimode USC conditions to explore the modulation of photoexcited carrier mobility through electron--phonon interactions in perovskite solar cells.

More broadly, our findings open new directions for phonon-based quantum technologies~\cite{Bienfait2019,Zivari2022,Qiao2023} in nonequilibrium scenarios, with applications ranging from the control of superconductivity~\cite{Babadi2017} and multimode entanglement~\cite{Andersson2022} to the generation of coherent THz sources~\cite{Greffet2002} and enhanced energy transfer~\cite{Maire2017} in solid-state systems. Notably, while optical phonons typically do not contribute to heat transfer due to their vanishing group velocity, in our system, the lower polariton (LP) acquires a finite group velocity and retains a substantial ($\sim 20\%$) phonon weight in the low-cavity-frequency regime. This suggests the intriguing possibility that superthermal phonon bunching in the multimode USC regime may influence heat transport in perovskite materials.
 
\bibliography{scibib}

\clearpage
\begin{figure}[t]
\centering
\includegraphics[width=0.7\linewidth]{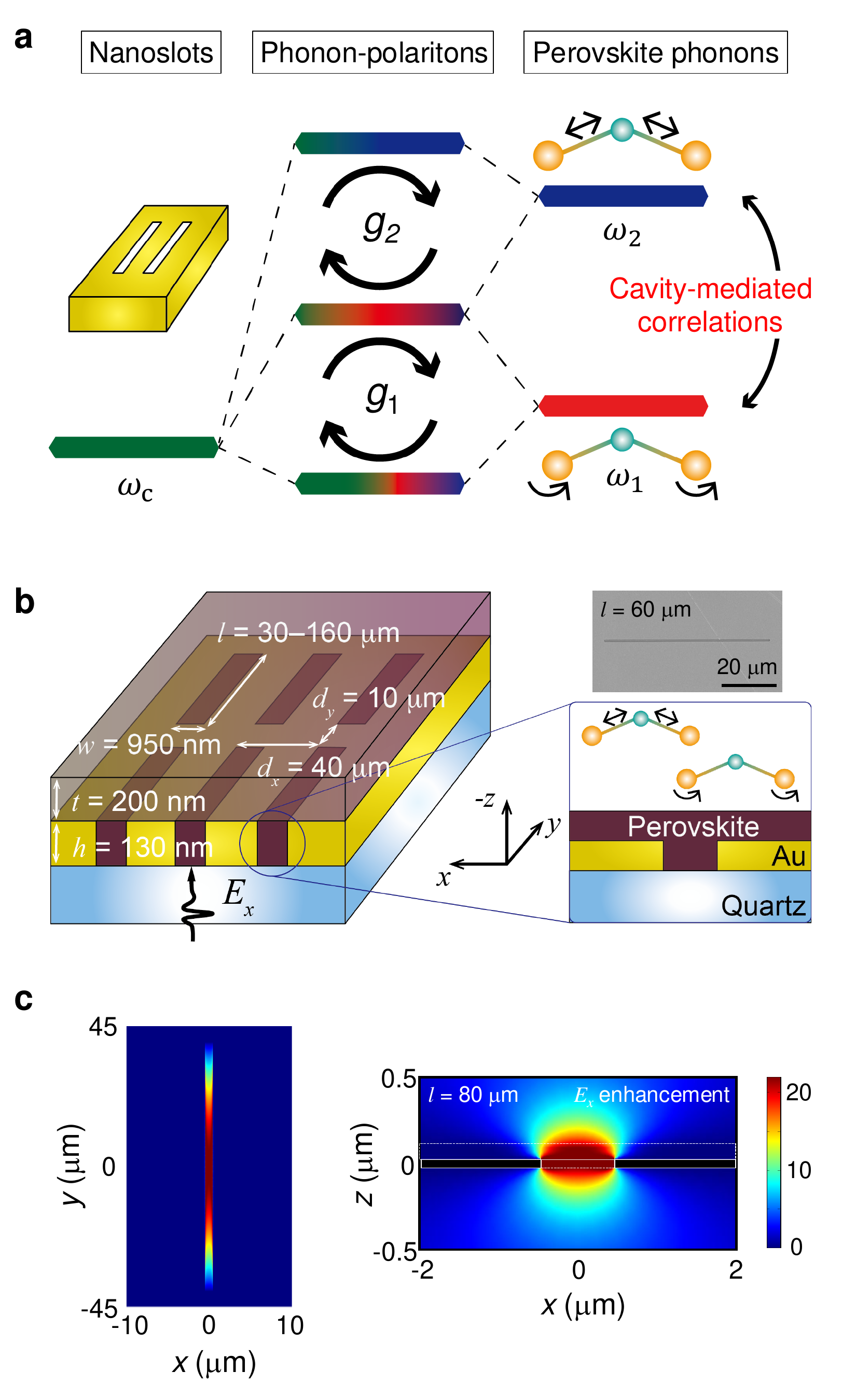}
\caption{\textbf{Perovskite--nanoslot hybrid system in the ultrastrong coupling regime.} \textbf{a},~Hybridization between a nanoslot-cavity mode, with frequency $\omega_\mathrm{c}$, and two transverse optical phonon modes in perovskite materials, with frequencies $\omega_{1}$ and $\omega_{2}$, in the far-detuned, low cavity frequency regime, $\omega_\mathrm{c}\ll \omega_{\lambda}$ ($\lambda=1,2$). The coupling strengths of these phonon modes are denoted as $g_1$ and $g_2$, respectively. Anomalous correlations between phonons are mediated by the cavity mode and governed by the coupling ratios $g_{\lambda}/\omega_{\lambda}$. \textbf{b},~Illustration of the perovskite--nanoslot hybrid system under illumination by terahertz light. Seven nanoslots of different lengths ($l=$ 30-160\,$\upmu$m) were fabricated to tune the cavity resonance frequency. The inset shows a scanning electron microscope image showing a bare nanoslot (top view); Scale bar: 20\,$\upmu$m. \textbf{c},~Numerical simulation (COMSOL) showing an enhancement of the $x$ component of the electric field ($E_x$) at resonance (0.77~THz) in a nanoslot filled with MAPbI$_3$ perovskite. Left: top view ($z=0$ plane); Right: cross-section ($y=0$ plane). The white dotted lines outline the area filled with MAPbI$_3$. The white solid lines outline the nanoslot area.}
\label{Fig1_main}
\end{figure}

\clearpage
\begin{figure}[t]
\centering
\includegraphics[width=1\linewidth]{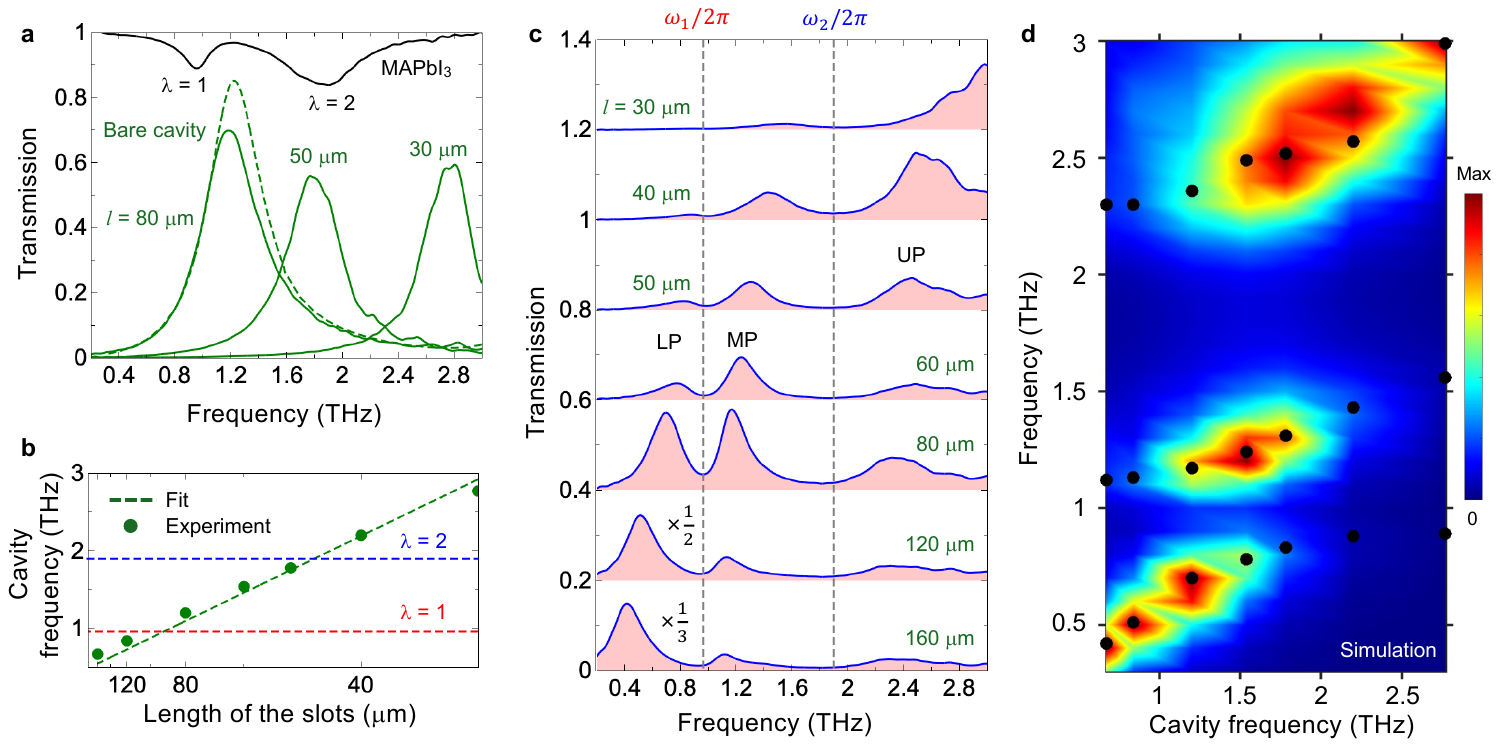}
\caption{\textbf{Terahertz transmission spectra.} \textbf{a},~Transmission spectra for bare cavities (nanoslots) with different lengths ($l$) (green curves) showing a single cavity mode. The green dashed line shows the simulated transmission through the nanoslot ($l=80\,\upmu$m). Transmission spectrum for a 200-nm-thick bare MAPbI$_3$ film (black curve) showing two transmission dips due to the two optical phonon modes ($\lambda=1$ and $\lambda=2$) with angular frequencies $\omega_{1}$ and $\omega_{2}$, respectively. \textbf{b},~Bare cavity resonance frequencies as a function of nanoslot length $l$ in the reciprocal axis (green circles). The linear fit (green dashed line) shows good agreement with the experimental data. The $\lambda=1$--cavity and $\lambda=2$--cavity resonances occur with an 80-$\upmu$m-long slot and with a 50-$\upmu$m-long slot, respectively, when the cavity mode frequency coincides with the phonon frequencies (red and blue dashed lines). \textbf{c},~Transmission spectra for the MAPbI$_3$--nanoslots hybrid system showing three polariton branches.  UP: upper polariton, MP: middle polariton, and LP: lower polariton. The dashed lines indicate the two phonon frequencies. The spectra are vertically offset by 0.2 for clarity. \textbf{d},~Numerical simulation (COMSOL) of the transmission as a function of cavity frequency (color map). Each spectrum has been normalized by its maximum transmittance to clearly show the three polariton branches; the black solid circles are the experimental results.}
\label{Fig2_main}
\end{figure}

\clearpage
\begin{figure}[t]
\centering
\includegraphics[width=1\linewidth]{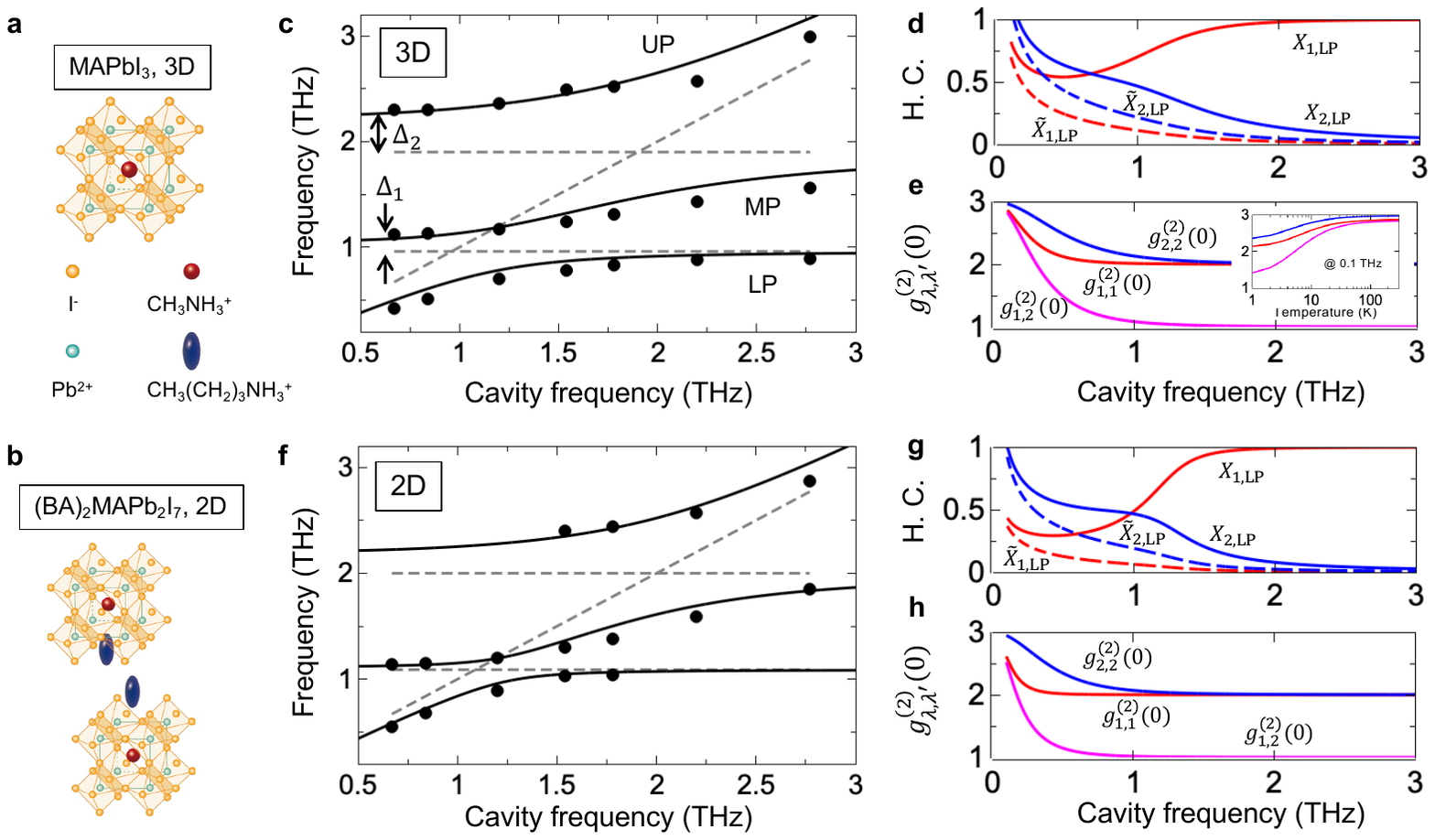}
\caption{\textbf{Phonon-polariton properties in perovskite--nanoslot hybrid systems.} 
Top: MAPbI$_3$ films (3D perovskite). Bottom: (BA)$_2$(MA)Pb$_2$I$_7$ (2D perovskite) films. 
\textbf{a},\textbf{b},~Crystal structures of MAPbI$_3$ and (BA)$_2$(MA)Pb$_2$I$_7$. BA: CH$_3$(CH$_2$)$_3$NH$_3^+$, MA: CH$_3$NH$_3^+$. \textbf{c},\textbf{f},~Polariton dispersion as a function of cavity frequency; UP: upper polariton, MP: middle polariton, LP: lower polariton. Solid circles: Peak frequencies extracted from the experimental transmission spectra. Solid lines: Fit of the extracted peak frequencies using the microscopic Hopfield model. The dashed lines indicate the $\lambda=1$ and $\lambda=2$ phonon modes and the cavity resonance. The two polariton gaps (see text) are denoted as $\Delta_{1}$ and $\Delta_{2}$. \textbf{d},\textbf{g},~Phonon Hopfield coefficients (H.C.) of the LP as a function of cavity frequency, showing a divergence in the low cavity frequency limit. \textbf{e},\textbf{h},~Theoretical predictions: Equal-time second-order phonon--phonon correlation functions $g^{(2)}_{\lambda,\lambda'} (\tau=0)$ for a polariton thermal state at room temperature as a function of cavity frequency. The inset in (\textbf{e}) shows $g^{(2)}_{\lambda,\lambda'} (0)$ as a function of temperature $T$ for a cavity frequency of 0.1~THz.}
\label{Fig3_main}
\end{figure}

\clearpage

\section*{Methods}
\subsection*{Sample preparation}\,\\
Dimethyl Formamide (DMF), Dimethyl Sulfoxide (DMSO), Lead Oxide (PbO), Butylamine (BA), Hydriodic acid (HI), Hypophosphorous acid (H$_3$PO$_2$) and Diethyl Ether were purchased from Sigma Aldrich and used without any further treatment. Methylammonium Iodide (MAI) and Methylammonium Chloride (MACl) were purchased from Greatcell Solar. Lead Iodide (PbI$_2$) was purchased from TCI Chemicals.

For the synthesis of MAPbI$_3$ (3D) films, the precursor solution was made by dissolving 95.4\,mg MAI, 276.6\,mg PbI$_2$ in 638\,$\mu$l DMF and 71\,$\mu$l DMSO and stirred on the hotplate at 70$^{\circ}$C for 3 hours. 4\,mg MACl was added to improve the film crystallinity. 70\,$\mu$l solution was then spin-coated on the nanoslots at 5000\,rpm and 3500\,rpm$/$s acceleration for 30 seconds. 600\,$\mu$l of Diethyl ether was dripped at 10 seconds from the start. The films were then annealed at 100$^{\circ}$C for 10 minutes.

(BA)$_2$(MA)Pb$_2$I$_7$ crystals (in the form of small plates) were prepared by adopting the previously reported procedure~\cite{Hou2024}. (BA)$_2$(MA)Pb$_2$I$_7$ (2D) films were made with the phase selective method as described in an earlier work~\cite{Sidhik2021}. Nominally, the parent crystals were dissolved in DMF at a concentration of 0.2\,M (30\,mg/100\,$\mu$l) and stirred on the hotplate at 70$^{\circ}$C for 2 hours in an argon glovebox. The solution was then transferred to a different glovebox where it was spin-coated on the nanoslots at 5000\,rpm with 3500\,rpm$/$s acceleration for 30 seconds. The films turn red-brown during the spin-coating process and are annealed at 100$^{\circ}$C for 5 minutes.

To fabricate the nanoslots, we utilized a standard photolithography technique to pattern photoresists to be an array of rods (950\,nm by $l$), followed by Au deposition (150\,nm) by an electron beam evaporator. Then, we performed an Ar beam ion milling on the samples to facilitate a lift-off process. In this process, the thickness of the Au films decreased to 130\,nm by the ion milling. After the lift-off process with acetone, we obtained an array of nanoslots. An array of fabricated bare nanoslots is presented in a scanning electron microscope image (top view) in the inset of Fig.~\ref{Fig1_main}b. Then, perovskite polycrystalline films ($\sim 200$~nm thick) were coated on the nanoslots.

\subsection*{THz time-domain spectroscopy (THz-TDS)}\,\\
We performed THz-TDS transmission measurements in a dry air environment at room temperature. The total measurement time for each sample was less than 15 minutes to avoid the degradation of perovskite films. To access high-frequency THz emission (up to 3\,THz), we utilized InGaAs photoconductive antennas for both emitter and detector which are fiber-coupled with an Er-fiber laser (100\,MHz, 1.5\,$\mu$m). Electric field amplitudes were low enough to avoid any field strength-dependent nonlinear effects. The emitted THz waves are guided to be sequentially focused on the samples and the detector by four 90\textdegree-off axis parabolic mirrors. The THz beam size at the focal point was about 1\,mm. To obtain transmission spectra of samples $\Tilde{T}=\left| E_\text{sample}(\omega)/E_\text{ref}(\omega) \right|^2$, we first measured transmitted electric fields $E_\text{sample}(t)$ of a sample and those of a bare quartz substrate $E_\text{ref}(t)$ as a reference, where $t$ is a delay time. Then, we performed the Fourier transformation to obtain $E_\text{sample}(\omega)$ and $E_\text{ref}(\omega)$.

\newpage
\section*{Acknowledgements}
D.K., A.B., F.T., and J.K.\ acknowledge support from the U.S.\ Army Research Office (through Award No.\ W911NF2110157), the W.\ M.\ Keck Foundation (through Award No.\ 995764), the Gordon and Betty Moore Foundation (through Grant No.\ 11520), and the Robert A.\ Welch Foundation (through Grant No.\ C-1509). W.W.\ and S.H.\ acknowledge support from the Air Force Office of Scientific Research (grant FA9550-22-1-0408), the National Science Foundation (Grant Nos.\ ECCS-2246564 and ECCS-1943895), and the Welch Foundation (Grant No.\ C-2144). E.E.M.C.\ acknowledges support from the Singapore Ministry of Education (MOE) Academic Research Fund Tier 3 grant (MOE-MOET32023-0003) ``Quantum Geometric Advantage.''

\section*{Author contributions} J.K.\ supervised the project. D.K.\ conceived the project, built the THz setup, performed all THz measurements and numerical simulations, analyzed experimental data, and prepared the manuscript under the supervision and guidance of J.K. D.H.\ developed the theoretical model, performed all calculations and fitting, and wrote the manuscript together with D.K. J.H.\ and A.A.\ grew the perovskite films under the guidance of A.D.M. G.L., S.K., and D.K.\ designed and fabricated the nanoslots under the guidance of M.S.\ and D.-S.K. All authors discussed the results and commented on the manuscript.

\section*{Competing interests} 
The authors declare no competing interests.

\section*{Additional information} Supplementary information is available. Correspondence and requests for materials should be addressed to J.K.

\section*{Data availability}
Data supporting this study's findings are available from the corresponding author upon reasonable request.

\section*{Code availability}
Codes supporting this study's findings are available from the corresponding author upon reasonable request.

\end{document}